\documentclass[
   ,final            
    ]
    {aipproc}

\usepackage{amsmath}
\usepackage{amssymb}

\layoutstyle{6x9}

\input{aipcheck.tex}

\begin{document}
	    
\title{
\vskip -2cm
\rightline{\parbox{3 cm}{\small \rm DESY 06-198\\HU-EP-06/38\\LMU-ACS 78/06\\MKPH-T-06-17}}
\vskip 1cm
Vacuum structure as seen by overlap fermions}


\author{E.-M. Ilgenfritz}{
 address={Humboldt-Universit\"at zu Berlin, Institut f\"ur Physik, 12489 Berlin, Germany}
}

\author{K. Koller}{
 address={Sektion Physik, Universit\"at M\"unchen, 80333 M\"unchen, Germany}
}

\author{Y. Koma}{
 address={Institut f\"ur Kernphysik, Johannes-Gutenberg Universit\"at Mainz,
 55099 Mainz, Germany}
}

\author{G. Schierholz}{
 address={Deutsches Elektronen-Synchrotron DESY, 22603 Hamburg, Germany}
 ,altaddress={John von Neumann-Institut f\"ur Computing NIC, 15738 Zeuthen, Germany}
}

\author{T. Streuer}{
 address={Department of Physics and Astronomy, University of Kentucky, Lexington, KY 40506-0055, USA}
}

\author{V. Weinberg}{
 address={Deutsches Elektronen-Synchrotron DESY, 15738 Zeuthen, Germany}
}

\begin{abstract}
Three complementary views on the QCD vacuum structure, 
all based on eigenmodes of the overlap operator, 
are reported in their interrelation: 
(i) spectral density, localization and chiral properties of the modes, 
(ii) the possibility of filtering the field strength with 
the aim to detect selfdual and antiselfdual domains and 
(iii) the various faces of the topological charge density, 
with and without a cutoff $\lambda_{\rm cut} = O(\Lambda_{QCD})$. 
The techniques are tested on quenched $SU(3)$ configurations. 
\end{abstract}

\maketitle


\vspace*{-0.5cm}
\section{Introduction}
\vspace*{-0.2cm}
Attempts to catch the structure of the QCD vacuum on the lattice 
were mostly inspired by the instanton/caloron picture, and appropriate 
lumps were searched for in the gluonic lattice topological charge density $q(x)$. 
With the advent of overlap fermions, the fermionic definition of the topological 
charge density has become viable, from the scale of the lattice 
spacing $a$ up to the infrared. This is possible since the Ginsparg-Wilson 
approach provides lattice fermions with perfect chiral properties.
Results concerning the vacuum structure, based on the eigenmodes of the overlap 
Dirac operator that have been collected in the QCDSF 
collaboration~\cite{in_preparation,Weinberg:2006ju}, 
are described in this contribution. 
\vspace*{-0.3cm}
\section{Properties of the lowest eigenmodes}
\vspace*{-0.2cm}
We analyze a large set of overlap eigenmodes, 
$O(150)$ per lattice configuration, for five ensembles of quenched 
Yang-Mills theory generated with the L\"uscher-Weisz action. This 
action is important to avoid dislocations. 
The analysis is performed on $12^3 \times 24$, $16^3 \times 32$ and $24^3 \times 48$ 
lattices at $\beta=8.45$, 
on a $12^3 \times 24$ lattice at $\beta=8.10$
and on a $16^3 \times 32$ lattice at $\beta=8.00$. 
In Fig. 1 (left) the spectral densities for the three volumes at $\beta=8.45$ are
presented. The fit to quenched chiral perturbation theory~\cite{Damgaard:2001xr} 
requires to know the distribution $w(Q)$ of topological charge,
easily obtained from the number and chirality $\pm1$ of the zeromodes, 
$Q=N_{-}-N_{+}$, per configuration. Fig. 1 shows the fit of the spectral 
densities according to
\begin{equation}
\rho(\lambda,V)=\Sigma_{\rm eff}(V,\lambda)\sum_Q w(Q) \rho_Q(\lambda~V~\Sigma_{\rm eff}(V,\lambda)) \; ,
\label{eq:rho}
\end{equation}
with the microscopic spectral density 
$\rho_Q(x)=(x/2)(J_{|Q|}^2(x)-J_{|Q|+1}(x)J_{|Q|-1}(x))$ in the fixed $Q$ sector,
leading to an estimate of the quark condensate, 
$\Sigma^{\overline{\mbox{\tiny MS}}}(\mbox{2\,GeV})=(231(20)\, \mbox{MeV})^3$.
\begin{figure}[!t]
\begin{tabular}{cc}
\includegraphics[width=6.5cm]{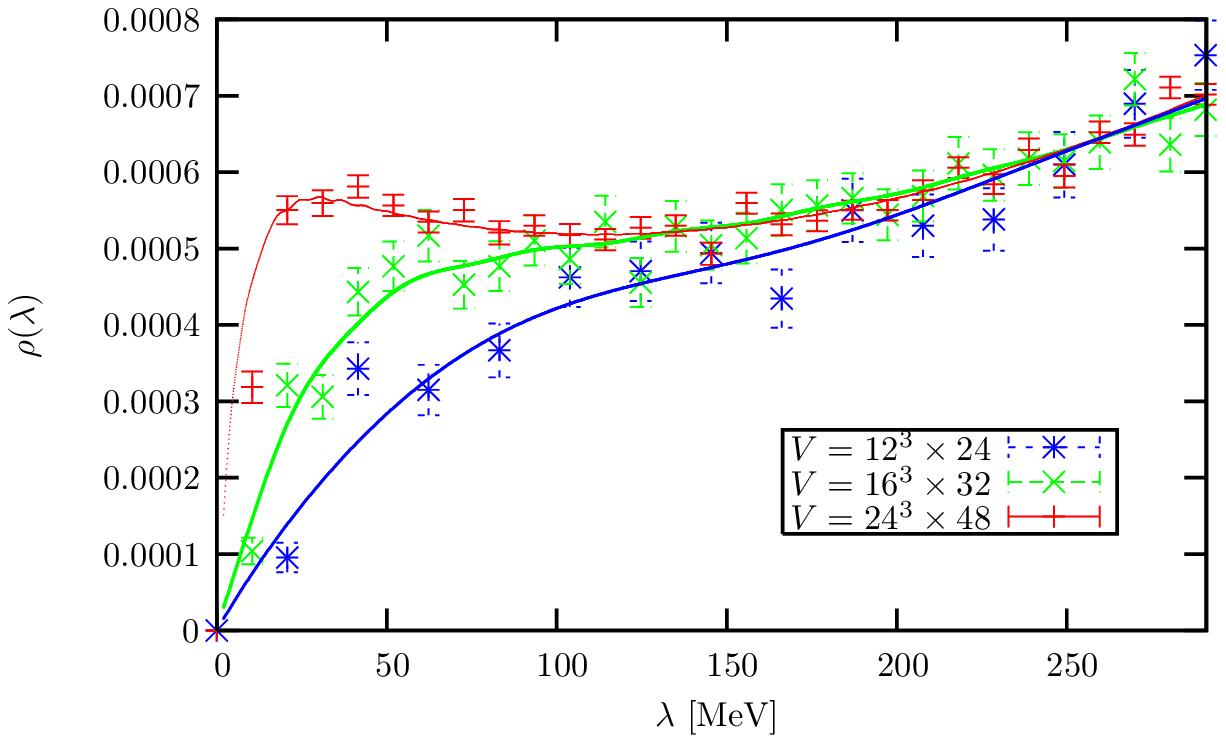}&
\hspace*{0.5cm}
\includegraphics[width=7.3cm]{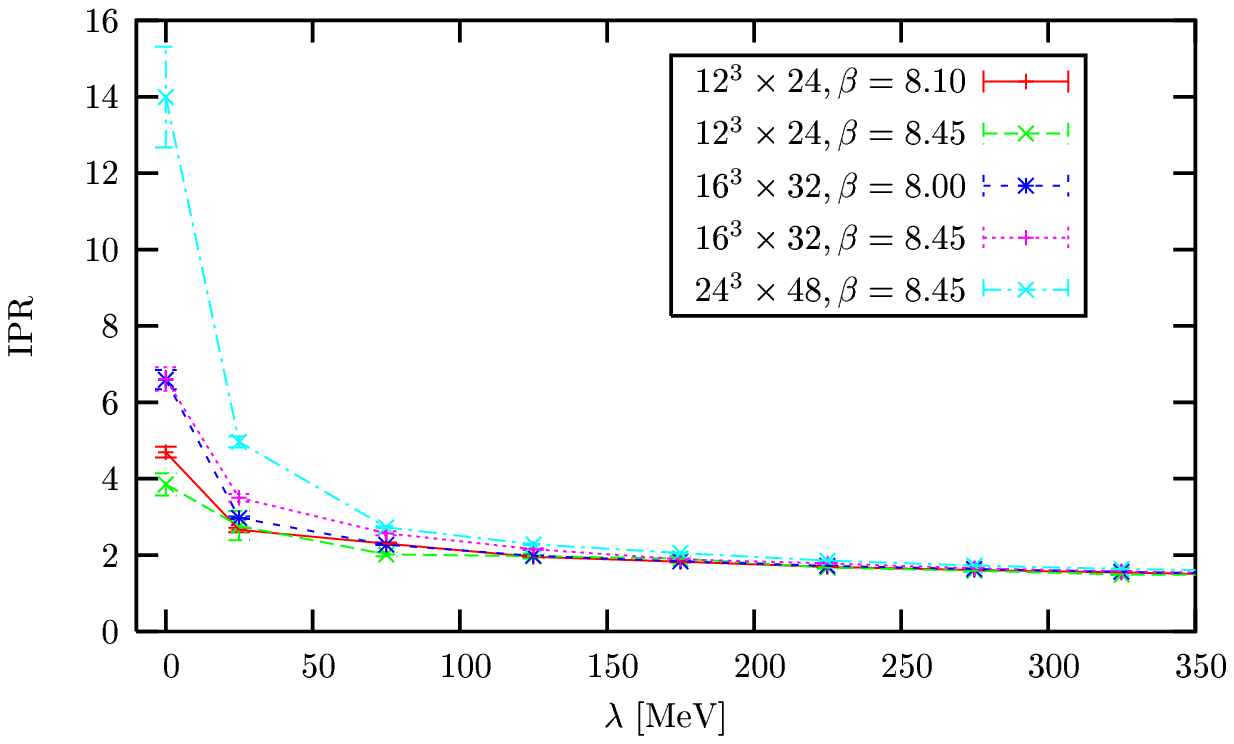} \\
\end{tabular}
\caption{Left: The spectral density of non-zeromodes for three lattice sizes 
at $\beta=8.45$, together with a simultaneous fit using the finite volume 
prediction of quenched chiral perturbation theory.
Right: The average IPR for zeromodes and for non-zeromodes in 
$\Delta\lambda=50$ MeV bins for all five ensembles.}
\label{fig:spec-ipr}
\end{figure}

In Fig. 1 (right) the average inverse participation ratio (IPR) of the zeromodes 
and non-zeromodes is presented. 
In the spectral region of strong volume dependence the modes are highly localized,
the spread of the IPRs is large and the level compressibility 
$\alpha = \langle (N - \langle N \rangle)^2 \rangle/\langle N \rangle << 1$
(for $N$ modes falling in some interval)
is small. In the theory of the metal-insulator transition this situation is called 
``critical'', and the modes are multifractal.
Fitting the $V$ dependence of the IPRs, one can assign a dimension 
$d^{*}(p=2) \approx 2$ to the zeromodes and $d^{*}(p=2) \approx 3 - 3.5$ to the 
non-zeromodes below 100 MeV (see Fig. 2 (left), upper curve). 
Similar results of the MILC collaboration~\cite{Aubin:2004mp} have been
obtained with Asqtad fermions. The IPRs of the higher modes 
are consistent with $d^{*}(2) \approx 4$. 
Generalized IPRs based on higher moments 
$I^{(p)}_{\lambda} = \sum_x p_{\lambda}(x)^p$ of the scalar density
$p_{\lambda}(x)=\psi^{\dagger}_{\lambda}(x)\psi_{\lambda}(x)$
localize the peaks of $p_{\lambda}(x)$ in regions of lower dimension, 
$d^{*}(p>2) < d^{*}(p=2)$. 
We find $d^{*}(p=10) < 1$ for zeromodes and $d^{*}(p=10) \sim 1$ 
for non-zeromodes with $\lambda < 100$ MeV. The multifractality evidenced
in Fig. 2 (left) urges a more detailed study of the lowest eigenmodes, 
in particular closer to the continuum limit. 
Case studies~\cite{Gattnar:2004gx} show that they might be 
pinned-down on low-dimension defects of the gauge field 
(monopoles and vortices, the candidates to create confinement).
\begin{figure}[!t]
\begin{tabular}{cc}
\includegraphics[width=7.3cm]{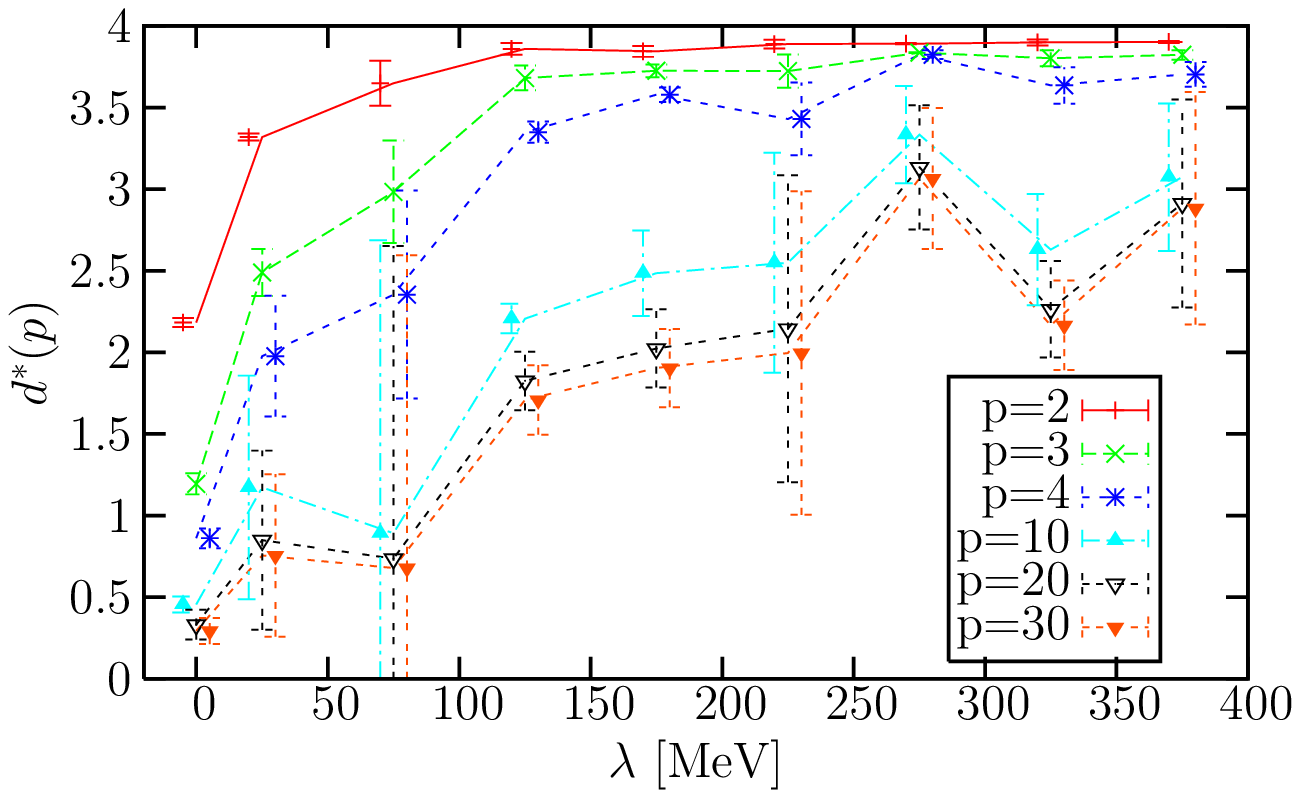}&
\includegraphics[width=6.5cm,height=4.5cm]{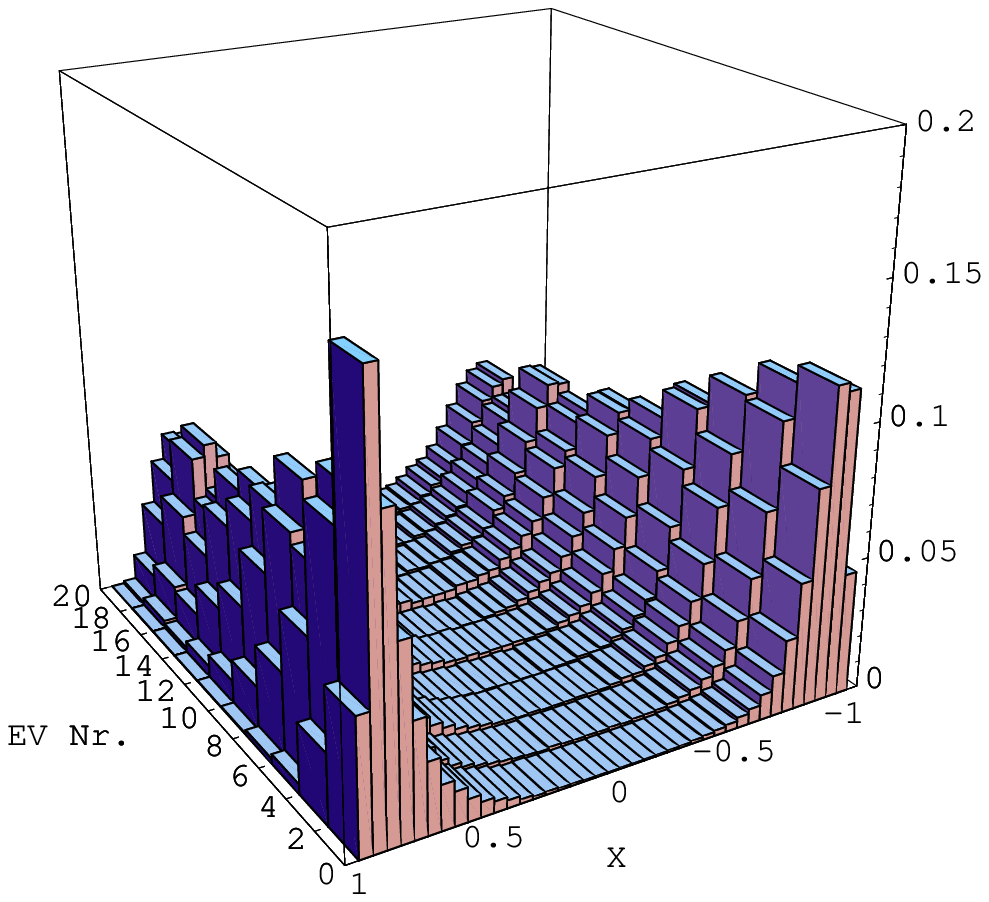} \\
\end{tabular}
\caption{Left: The multifractal dimensions of zeromodes and of non-zeromodes, 
from the volume dependence of average $I^{(p)}_{\lambda}$ for 
the three ensembles at $\beta=8.45$. Right: The histograms of local chirality 
$X_{\lambda}(x)$ for the 10 lowest pairs of non-zeromodes over 1 \% of sites 
with highest scalar density.} 
\label{fig:dim-localchir}
\end{figure}

Different from the high temperature phase~\cite{DIK}, the distribution of local 
chirality of the lowest non-zeromodes peaks at non-zero values of the 
chirality variable
\begin{equation}
X_{\lambda}(x) = (4/\pi)~\arctan \sqrt{p_{\lambda+}(x)/p_{\lambda-}(x)} - 1 \in [-1,+1] \; .
\label{eq:Xarctan}
\end{equation}
expressing the ratio between the density of the chirality components, 
$p_{\lambda\pm}(x)=\frac{1}{2}\psi^{\dagger}_{\lambda}(x)(1 \pm \gamma_5)\psi_{\lambda}(x)$.
Focussing on the highest-ranking 1 \% of lattice sites according to the scalar 
density $p_{\lambda}(x)$, 
the histograms for the 10 lowest pairs $\pm \lambda$ are shown in Fig. 2 (right).

\vspace*{-0.3cm}
\section{Topological density and a measure of selfduality}
\vspace*{-0.2cm}
The {\it all-scale} topological charge density in terms of the overlap operator 
$D_N$ is
\begin{equation}
q(x)=-~\mathrm{Tr}_{color,spinor}~\left[ \gamma_5~\left( 1 - \frac{a}{2}~D_N(x,x) \right) \right] \; , 
\label{eq:def-q}
\end{equation}
while the UV-filtered version $q_{\lambda_{\rm cut}}(x)$ gets contributions to 
the trace only from modes with $|\lambda| < \lambda_{\rm cut} \sim \Lambda_{QCD}$.

The density $q(x)$ satisfies the negativity of the topological 
correlator $C_q(r)$~\cite{Koma:2005sw} 
for $r \ne 0$
due to the diverging multiplicity of clusters at large enough 
$\beta$ or $a \lesssim 0.1$ fm. At $\beta=8.45$ one finds $\approx 75$ such 
clusters per fm$^{4}$ that percolate at $|q(x)|=0.25~q_{\rm max}$. 
Clusters with different fractal dimensions $d^{*}<3$ are visible 
at $|q(x)|$ above the percolation threshold. The global 3D structure discovered by 
Horvath {\it et al.}~\cite{Horvath:2003yj} appears below that level.
The UV-filtered density with $\lambda_{\rm cut}=200$ MeV forms
only one cluster per fm$^{4}$ percolating at 
$|q_{\lambda_{\rm cut}}(x)|=0.05~q_{\lambda_{\rm cut}\rm max}$.

Gattringer~\cite{Gattringer:2002gn} proposed an UV filter for the field strength 
tensor using eigenmodes $\psi_{\lambda}(x)$ of
the Dirac operator ($T^a$ denotes a color generator in the fundamental 
representation)  
\begin{equation}
F^a_{\mu\nu}(x) \propto \sum_j \lambda^2_j f^a_{\mu\nu}(x|j) \; , \mathrm{~with~}   
f^a_{\mu\nu}(x|j) = -(i/2) \psi^{\dagger}_{\lambda_j}(x) 
\gamma_{\mu} \gamma_{\nu} T^a \psi_{\lambda_j}(x) \; .
\end{equation}
This prescription yields a filtered field strength up to an (undetermined) 
normalization when only low-lying modes are included. 
With a filtered action density
$\tilde{s}(x)=\mathrm{Tr}~F_{\mu\nu}(x) F_{\mu\nu}(x)$ and topological charge density
$\tilde{q}(x)=\mathrm{Tr}~F_{\mu\nu}(x) \tilde{F}_{\mu\nu}(x)$,  
one gets an estimator for the local (anti-)selfduality analogous to
Eq. (\ref{eq:Xarctan}), 
\begin{equation}
R(x) = (4/\pi)~\arctan \sqrt{(\tilde{s}(x)-\tilde{q}(x))/
(\tilde{s}(x)+\tilde{q}(x))} - 1 \in [-1,+1] \; .
\label{eq:Rarctan}
\end{equation}
This observable has good cluster properties if a cut separates 
(anti-)selfdual domains from the rest of the lattice. 
For $R_{\rm cut} \lesssim 1$ the number of such domains 
is practically stable under changing the cutoff. 
It depends, however, on the number of modes used in the filter. 
$R(x)$ is correlated with the local chirality $X_{\lambda}(x)\approx \pm 1$ 
of the low eigenmodes, and the $R$-cluster structure corresponds to the 
cluster structure of $q_{\lambda_{\rm cut}}$, as demonstrated in Fig. 3,
if both are tuned to the relevant number of modes 
($\lambda_{\rm cut} \sim \Lambda_{QCD}$). 
\begin{figure}[!t]   
\begin{tabular}{cc}
\includegraphics[width=4.8cm]{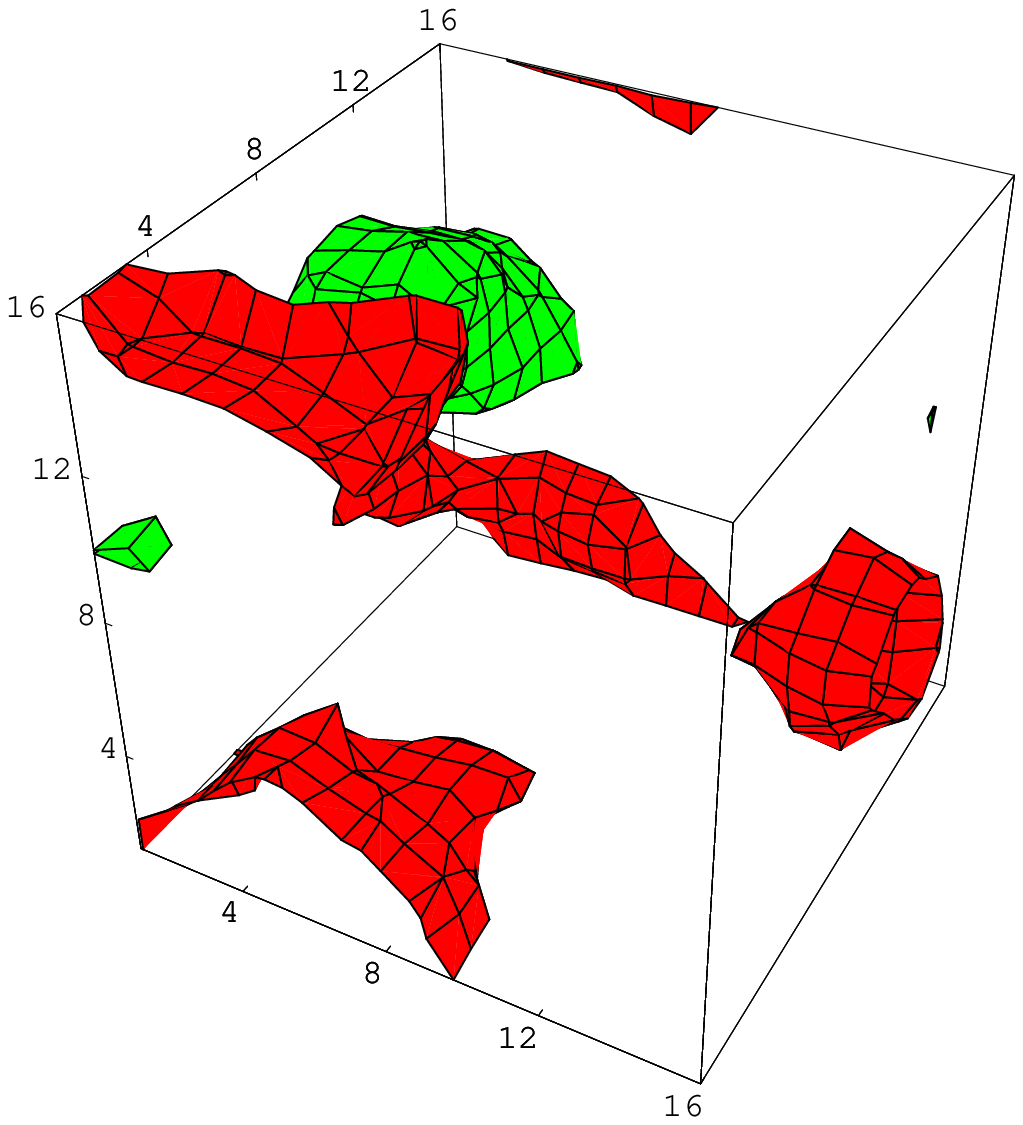}&
\hspace*{0.5cm}
\includegraphics[width=4.8cm]{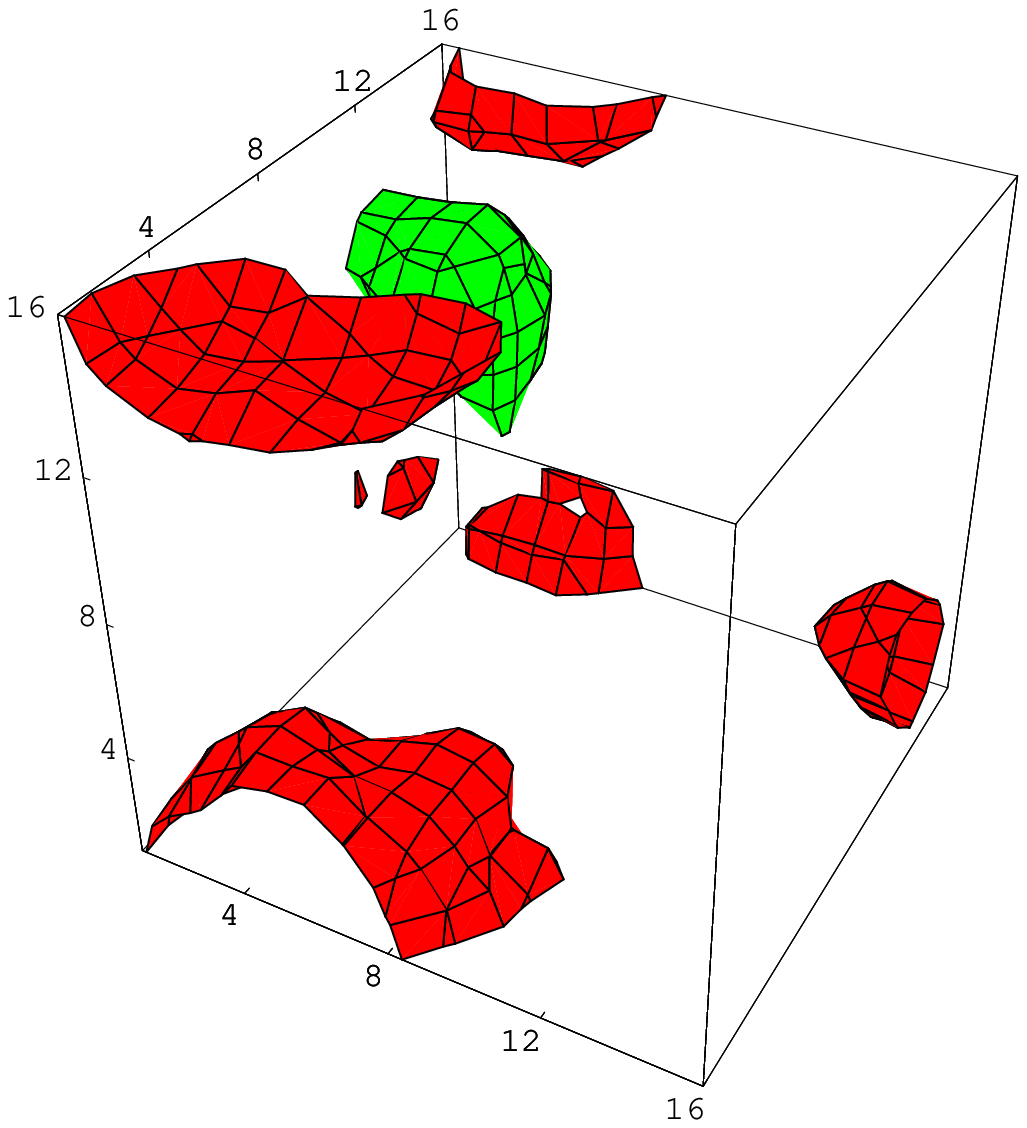}\\
\end{tabular}
\caption{Comparison of the $R$-clusters for $R_{\rm cut}=0.99$ with the
$q_{\lambda_{\rm cut}}$-clusters for $\lambda_{\rm cut}=100$ MeV 
(at 1/10 of the maximal density)
for one timeslice of a $16^3 \times 32$ lattice configuration with $Q=0$ 
generated at $\beta=8.45$.  The similarity is the same in all timeslices.}
\label{fig:compare-isosurfaces}
\end{figure}

\vspace*{-0.4cm}
\section{Conclusion}
\vspace*{-0.2cm}
The main conclusion is that overlap fermions can probe the lattice vacuum both 
in the ultraviolet and in the infrared. 
To see the phenomenologically known structure of 4D selfdual domains, some UV 
filtering by applying $\lambda_{\rm cut} \sim \Lambda_{QCD}$ is necessary 
without the need of smoothing the gauge field.
The {\it all-scale} topological charge density is multifractal, too, within
the global 3D structure. It forms lower dimensional clusters above the percolation 
threshold. 
The phenomenological significance of the low-dimensional (singular) structures 
is still speculative.


\vspace*{-0.4cm}
\begin{theacknowledgments}
\vspace*{-0.4cm}
E.-M.\ I.\ is grateful to the organizers for the opportunity to present this work.
He is supported by the DFG Forschergruppe FOR 465 
``Gitter-Hadronen-Ph\"anomenologie''. 
\end{theacknowledgments}
\vspace*{-0.3cm}



\bibliographystyle{aipproc}   

\vspace*{-0.2cm}
\bibliography{azoren2}
\vspace*{-0.2cm}

\IfFileExists{\jobname.bbl}{}
 {\typeout{}
  \typeout{******************************************}
  \typeout{** Please run "bibtex \jobname" to optain}
  \typeout{** the bibliography and then re-run LaTeX}
  \typeout{** twice to fix the references!}
  \typeout{******************************************}
  \typeout{}
 }

\end{document}